\documentclass[twocolumn]{bmcart}

\usepackage{amsthm,amsmath}
\usepackage{SIunits}
\usepackage[bookmarks=false]{hyperref}
\usepackage[capitalize]{cleveref}
\usepackage{graphicx}
\graphicspath{{figures/}}
\usepackage{tabularx}
\usepackage{multirow}
\usepackage{bm}
\usepackage{tikz}
\usepackage{natbib}
\usepackage[normalem]{ulem}

\definecolor{pink}{RGB}{255,0,255}

\definecolor{dark-greenish-turquoise}{RGB}{0,160,110}

\begin{document}

\begin{frontmatter}

\begin{fmbox}
\dochead{Research}

\title{Mitigating radiation damage of single photon detectors for space applications}

\author[
   addressref={aff1,aff2},
   corref={aff1,aff2},
   email={anisimovaa@gmail.com}
]{\inits{EA}\fnm{Elena} \snm{Anisimova}}
\author[
   addressref={aff1,aff2},
   email={brendon.higgins@uwaterloo.ca}
]{\inits{BH}\fnm{Brendon L} \snm{Higgins}}
\author[
   addressref={aff1,aff2},
   email={jbourgoin@uwaterloo.ca}
]{\inits{JPB}\fnm{Jean-Philippe} \snm{Bourgoin}}
\author[
   addressref={aff1},
   email={miles.cranmer@gmail.com}
]{\inits{MC}\fnm{Miles} \snm{Cranmer}}
\author[
   addressref={aff4},
   email={Eric.Choi@magellan.aero}
]{\inits{EC}\fnm{Eric} \snm{Choi}}
\author[
   addressref={aff3},
   email={danya.hudson@honeywell.com}
]{\inits{DH}\fnm{Danya} \snm{Hudson}}
\author[
   addressref={aff3},
   ]{\inits{LP}\fnm{Louis P} \snm{Piche}}
	\author[
   addressref={aff3},
   email={alan.scott@honeywell.com}
]{\inits{AS}\fnm{Alan} \snm{Scott}}
\author[
   addressref={aff2,aff1,aff5},
   email={makarov@vad1.com}
]{\inits{VM}\fnm{Vadim} \snm{Makarov}}
\author[
   addressref={aff1,aff2,aff6},
   email={thomas.jennewein@uwaterloo.ca}
]{\inits{TJ}\fnm{Thomas} \snm{Jennewein}}

\address[id=aff1]{
  \orgname{Institute for Quantum Computing, University of Waterloo},
  \city{Waterloo, ON},
	\postcode{N2L~3G1}
  \cny{Canada}
}
\address[id=aff2]{
  \orgname{Department of Physics and Astronomy, University of Waterloo},
  \city{Waterloo, ON},
  \postcode{N2L~3G1}
  \cny{Canada}
}
\address[id=aff3]{
  \orgname{Honeywell Aerospace (formerly COM~DEV Ltd.)},
  \street{303 Terry Fox Dr., Suite 100},
  \city{Ottawa, ON},
  \postcode{K2K~3J1}
  \cny{Canada}
}
\address[id=aff4]{
  \orgname{Magellan Aerospace},
  \street{3701 Carling Avenue},
  \city{Ottawa, ON},
  \postcode{K2H~8S2}
  \cny{Canada}
}
\address[id=aff5]{
  \orgname{Department of Electrical and Computer Engineering, University of Waterloo},
  \city{Waterloo, ON},
  \postcode{N2L~3G1}
  \cny{Canada}
}
\address[id=aff6]{
  \orgname{Quantum Information Science Program, Canadian Institute for Advanced Research},
  \city{Toronto, ON},
  \postcode{M5G~1Z8}
  \cny{Canada}
}

\begin{abstractbox}
\begin{abstract}
Single-photon detectors in space must retain useful performance characteristics despite being bombarded with sub-atomic particles. Mitigating the effects of this space radiation is vital to enabling new space applications which require high-fidelity single-photon detection. To this end, we conducted proton radiation tests of various models of avalanche photodiodes (APDs) and one model of photomultiplier tube potentially suitable for satellite-based quantum communications. The samples were irradiated with 106~MeV protons at doses approximately equivalent to lifetimes of 0.6, 6, 12, and 24~months in a low-Earth polar orbit. Although most detection properties were preserved, including efficiency, timing jitter and afterpulsing probability, all APD samples demonstrated significant increases in dark count rate (DCR) due to radiation-induced damage, many orders of magnitude higher than the 200 counts per second (cps) required for ground-to-satellite quantum communications. We then successfully demonstrated the mitigation of this DCR degradation through the use of deep cooling, to as low as $-86\,\celsius$. This achieved DCR below the required 200~cps over the 24 months orbit duration. DCR was further reduced by thermal annealing at temperatures of $+50$ to $+100\,\celsius$.
\end{abstract}


\end{abstractbox}
\end{fmbox}

\end{frontmatter}

\section*{Introduction}
\label{sec:intro}

Single-photon detectors (SPDs) have been utilized in a number of space applications, including laser ranging (LIDAR) for atmospheric and topology measurements of the Earth \cite{schutz2005, sun2004}, elementary particle scintillation detectors \cite{lecoq2011}, and precise laser time transfer \cite{prochazka2009}. SPDs will also be necessary to support quantum communication applications \cite{rarity2002, ursin2009, bonato2009, bourgoin2013, gibney2016, joshi2017}, where high detection efficiency, low timing jitter, low dark count rate (DCR) and low afterpulsing probability are key parameters for achieving successful, high-fidelity transmissions \cite{bourgoin2013, bourgoin2015}. Photomultiplier tubes (PMTs) and silicon avalanche photodiodes (APDs) are two types of SPDs that generally have good performance for this application, whereas superconducting nano-wire detectors may offer better performance, in some respects, at the cost of being significantly less practical, requiring cryogenic cooling \cite{hadfield2009}.

For optical transmissions through the atmosphere, a low-loss window exists at around 800~nm wavelength~\cite{bourgoin2013}. PMTs have reduced detection efficiencies for wavelengths longer than 650~nm, but silicon-based APDs have high detection efficiency in that region, low timing jitter, low DCR, and low afterpulsing, making them a prime candidate technology for quantum communication applications. However, incident radiation significantly increases the DCR of APDs \cite{sun1997, sun2001, tan2013, marisaldi2011, moscatelli2013}, which can quickly turn an APD unsuitable for quantum communications on a space platform.

Successful ground-to-satellite quantum communication requires each detector's DCR to be kept below about $200$ counts per second (cps)~\cite{bourgoin2013}. Previous use of silicon APD technology (specifically, Excelitas SLiK devices) for photon detection on a satellite showed an increase in dark count rates by ${\sim}30$~cps for each day in orbit~\cite{sun2004}, which would make them unusable for quantum communications in merely a few weeks. Other ground-based radiation tests of APDs also demonstrated DCRs too high for quantum communications \cite{sun1997, sun2001, marisaldi2011, moscatelli2013, tan2013}.

Recently reported tests attempted mitigation by cooling to temperatures as low as $-20\,\celsius$ to overcome the increased DCR~\cite{tan2013}. It is known that the DCR of non-irradiated APDs can be reduced by deeper cooling, decreasing the rate of thermally induced spontaneous avalanches \cite{kim2011}, but at the same time cooling increases the lifetimes of trapped carriers that contribute to afterpulsing, which may interfere with quantum communication~\cite{kim2011, anisimova2017a}. Thermal annealing was also found to reduce the DCR after irradiation~\cite{tan2013, sun2001, moscatelli2013}. However, no previously reported tests have applied deep cooling to radiation damaged APDs, nor have any demonstrated a sufficiently low DCR required for quantum communications, specifically quantum key distribution (QKD), or verified other detector parameters throughout a reasonable lifetime (e.g.,\ 1~year for an initial demonstrator mission) of a quantum receiver satellite.

Here we show experimentally that the effects of radiation doses approximately equivalent to as much as 2 years in low-Earth orbit are successfully mitigated by cooling and thermal annealing, allowing APDs to be used in a quantum satellite. We have tested three APD device models---Excelitas C30921SH and Laser Components SAP500S2 (each with sensitive areas {500\,\micro\meter} in diameter), and Excelitas SLiK (with sensitive area {180\,\micro\meter} in diameter)---and one PMT device model---Hamamatsu H7422P-40. All samples survived irradiation and remained functional photon detectors, with the only significant effect being the increase of the DCR in all APD samples. Breakdown voltage, afterpulsing, detection efficiency and timing jitter of the irradiated APDs were characterized and shown to be in the range acceptable for quantum communications. PMTs were also tested for dark counts, timing jitter, afterpulsing and detection efficiency.

The paper is structured as follows. In \nameref{sec:test} we describe how the equivalent radiation doses for devices under test were calculated, the design of our setup and the radiation test procedure. In \nameref{sec:effects} we present the measured radiation damage effects, and in \nameref{sec:mitigation} we demonstrate results of cooling and annealing on irradiated APDs. We finally give concluding remarks in \nameref{sec:conclusion}.

\section*{Radiation test} 
\label{sec:test}

\subsection*{Radiation dose and tested devices}

SPDs in low-Earth orbit experience space radiation primarily in the form of protons, electrons and heavy ions, resulting in two types of permanent damage in the semiconductor material: displacement and ionization damage \cite{messenger1986, srour1988, johnston2000}. APDs are less sensitive to ionization damage; e.g.,\ Ref.~\cite{tan2013} demonstrated that after 1-year equivalent ionization damage (in a 800~km equatorial orbit) Si APDs increased DCRs up to 2 times. However, displacement damage causes new defects in the semiconductor lattice of the active area, significantly affecting the DCR; e.g.,\ in Ref.~\cite{tan2013} DCR of APD irradiated by protons increased by one to two orders of magnitude (limited by a saturated passive quenching window comparator).

Dark current in APDs has two components: surface currents, which are unaffected by gain, and bulk leakage current which passes through the avalanche region and is therefore gain multiplied. Bulk dark current generation is linked directly to non-ionizing energy loss in a variety of silicon semiconductors \cite{srour2000}. Ionization damage is mainly associated with surface oxide interface dark current, and was not directly considered in this study. Afterpulsing is caused by delayed emission of trapped charge from bulk defects, in a thermally activated process (analogous to charge transfer efficiency losses in charge-coupled devices).

Proton displacement damage arises due to structural displacements in the silicon crystal caused by elastic collisions, and inelastic spallation reactions. The distribution of energies of trapped protons in low-Earth orbit, transported through 10~mm of aluminum shielding, possesses a broad peak in the range of 50 to 75~MeV. Here the ratio between elastic and inelastic energy loss ranges from 1.7 to 1.2, whereas at 100~MeV the ratio is roughly 1.0. Following a commonly accepted silicon damage deposition model \cite{dale1994}, we calculate the monochromatic proton fluence that produces the same average specific non-ionizing energy loss in silicon.

Due to this difference in the energy distribution ratio, the physical range of damage fragments through the sensitive microvolume of the detector will also be different, because inelastic reactions result in a much greater variance in the range of fragments in the silicon, compared to elastic damage which is uniformly distributed throughout. (That is, the damage energy equilibrium may not be established until several micrometers below the Si surface from the direction of incident proton flux.) This would result in under-dosing of the first few micrometers near the surface of the APD---at 100~MeV, damage equilibrium is not reached until about $3$ to $5~\micro\meter$ beneath the surface \cite{dale1994}. However, Ref.~\cite{becker2003}, which shows the internal structure of different types of APD, suggests that the important amplification region is typically tens of micrometers below the surface, where these small damage energy distribution differences will not be a major factor.

Following Ref.~\cite{bourgoin2013}, we chose a polar orbit at 600~km altitude, providing global coverage, as representative for our hypothetical quantum satellite. With a hypothetical shielding of 10-mm-thick aluminum around its detectors, predicted radiation doses were calculated using the SPENVIS radiation modeling tool for durations of 0.6, 6, 12, and 24~months. The radiation doses were determined to be equivalent to 100~MeV proton fluences of $10^8$, $10^9$, $2\times10^9$, and $4\times10^9~\centi\meter^{-2}$, respectively.

\begin{table}
\caption{\label{table:table_groups}Nine groups of tested samples and their corresponding nominal radiation fluences, equivalent to in-orbit exposures over 0.6, 6, 12, and 24~months with protons at $100~\mega\electronvolt$. Each APD in group 5 was biased during irradiation at $20~\volt$ above its breakdown voltage. Group 9 was not irradiated, and kept as a control.\vspace{1.4mm}}
\newlength{\colwidthdevice}
\newlength{\colwidthfluence}
\setlength{\colwidthdevice}{4cm}
\setlength{\colwidthfluence}{2cm}
\begin{tabular}{c|c|c} \hline \hline
 & & \\
 \textbf{Group} & \parbox[c]{\colwidthdevice}{\textbf{Device type and quantity}} & \parbox[c]{\colwidthfluence}{\textbf{Fluence\\@~100~MeV,\\protons/cm$^2$}} \\
 & & \\ \hline
 & & \\
1 & \parbox[c]{\colwidthdevice}{SLiK -- 2 pcs \\ SAP500S2 -- 2 pcs} & $10^8$ \\
\phantom{\vspace{3mm}} & & \\
2 & \parbox[c]{\colwidthdevice}{SLiK -- 2 pcs \\ SAP500S2 -- 2 pcs} & $10^9$ \\
\phantom{\vspace{3mm}} & & \\
3 & \parbox[c]{\colwidthdevice}{SLiK -- 2 pcs \\ SAP500S2 -- 2 pcs \\ C30921SH -- 2 pcs} & $2\times10^9$ \\
\phantom{\vspace{3mm}} & & \\
4 & \parbox[c]{\colwidthdevice}{SLiK -- 2 pcs \\ SAP500S2 -- 2 pcs \\ C30921SH -- 2 pcs} & $4\times10^9$ \\
\phantom{\vspace{3mm}} & & \\
5 & \parbox[c]{\colwidthdevice}{SLiK -- 2 pcs \\ SAP500S2 -- 2 pcs \\ C30921SH -- 2 pcs} & \parbox[c]{\colwidthfluence}{$4\times10^9$ (biased)} \\
\phantom{\vspace{3mm}} & & \\
6 & \parbox[c]{\colwidthdevice}{H7422-40 -- 1 pc} & $10^9$ \\
 & & \\
7 & \parbox[c]{\colwidthdevice}{H7422-40 -- 1 pc} & $2\times10^9$ \\
 & & \\
8 & \parbox[c]{\colwidthdevice}{H7422-40 -- 1 pc} & $4\times10^9$ \\
\phantom{\vspace{3mm}} & & \\
9 & \parbox[c]{\colwidthdevice}{SLiK -- 2 pcs \\ SAP500S2 -- 2 pcs \\ C30921SH -- 2 pcs \\ H7422-40 -- 1 pc} & 0 \\
 & & \\ \hline \hline
\end{tabular}
\end{table}

We tested a total of 32 APD devices and 4 PMT devices. These samples were divided among nine groups (see~\cref{table:table_groups}). We applied each of the four fluences to the first four groups with the devices switched off. For the fifth group, APD bias voltage was applied during irradiation at the highest fluence (24~month equivalent) to examine whether bias voltage affects the extent of damage caused by irradiation. The last group of samples was kept as a control group, being stored and transported alongside the other five groups, but without undergoing irradiation. The irradiation was done at the Tri-University Meson Facility (TRIUMF) at the University of British Columbia using a $106~\mega\electronvolt$ proton beam, which was slightly higher energy than the nominal $100~\mega\electronvolt$.

\subsection*{Characterization setup}

\begin{figure}[t]
  \includegraphics[width=0.94\columnwidth]{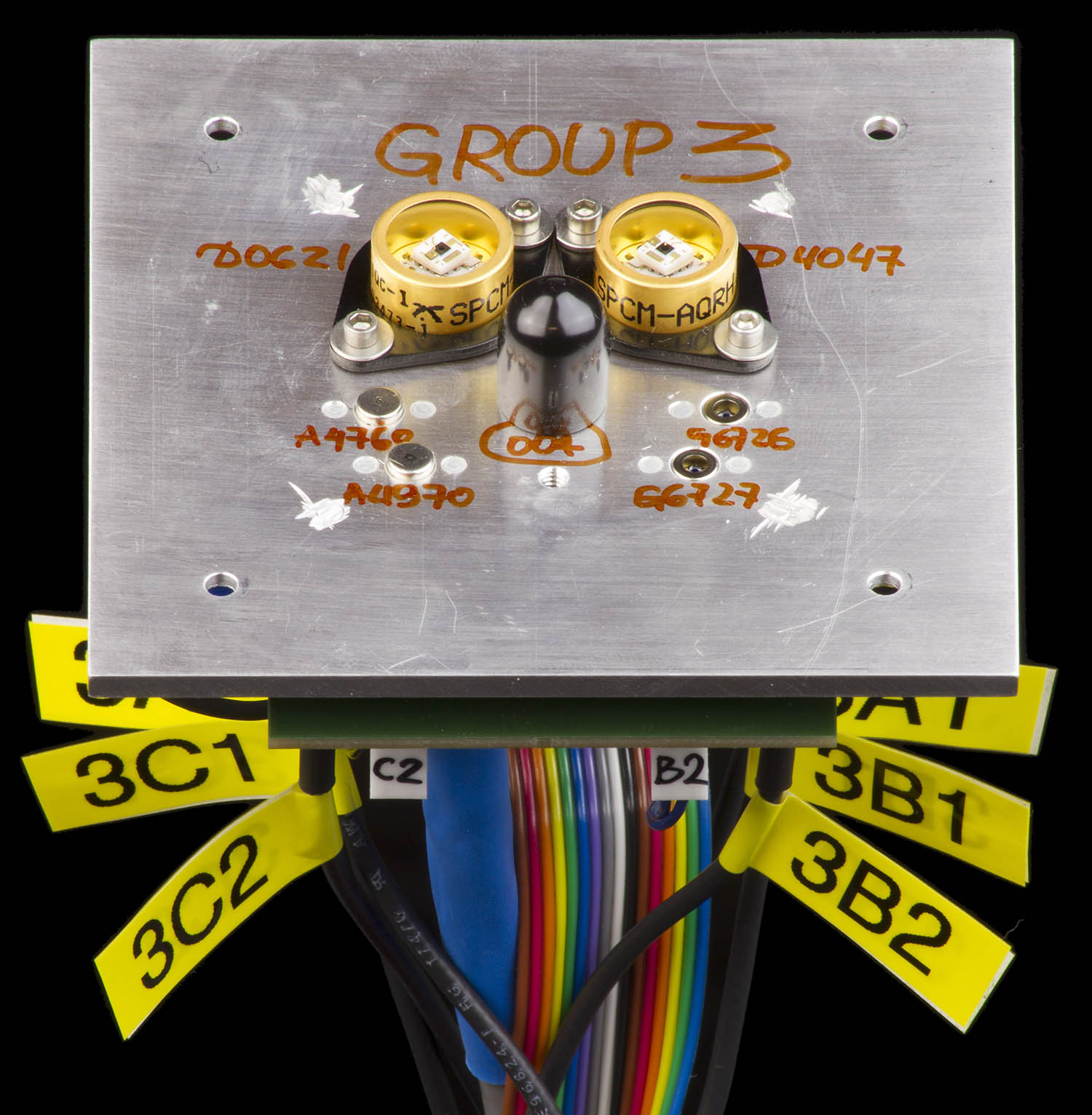}
  \caption{\label{fig:group_view}One group of APDs, consisting of two SLiK devices (top), two C30921SH devices (bottom left), and two SAP500S2 devices (bottom right). (The device under the black cap, center, is not discussed in this paper.) The detectors are connected to a PCB with 6 passive quenching circuits, attached to the back of the plate. Bias voltage supply and signal cables can be seen exiting from behind (far bottom).}
\end{figure}

\begin{figure}[t]
  \includegraphics[width=0.94\columnwidth]{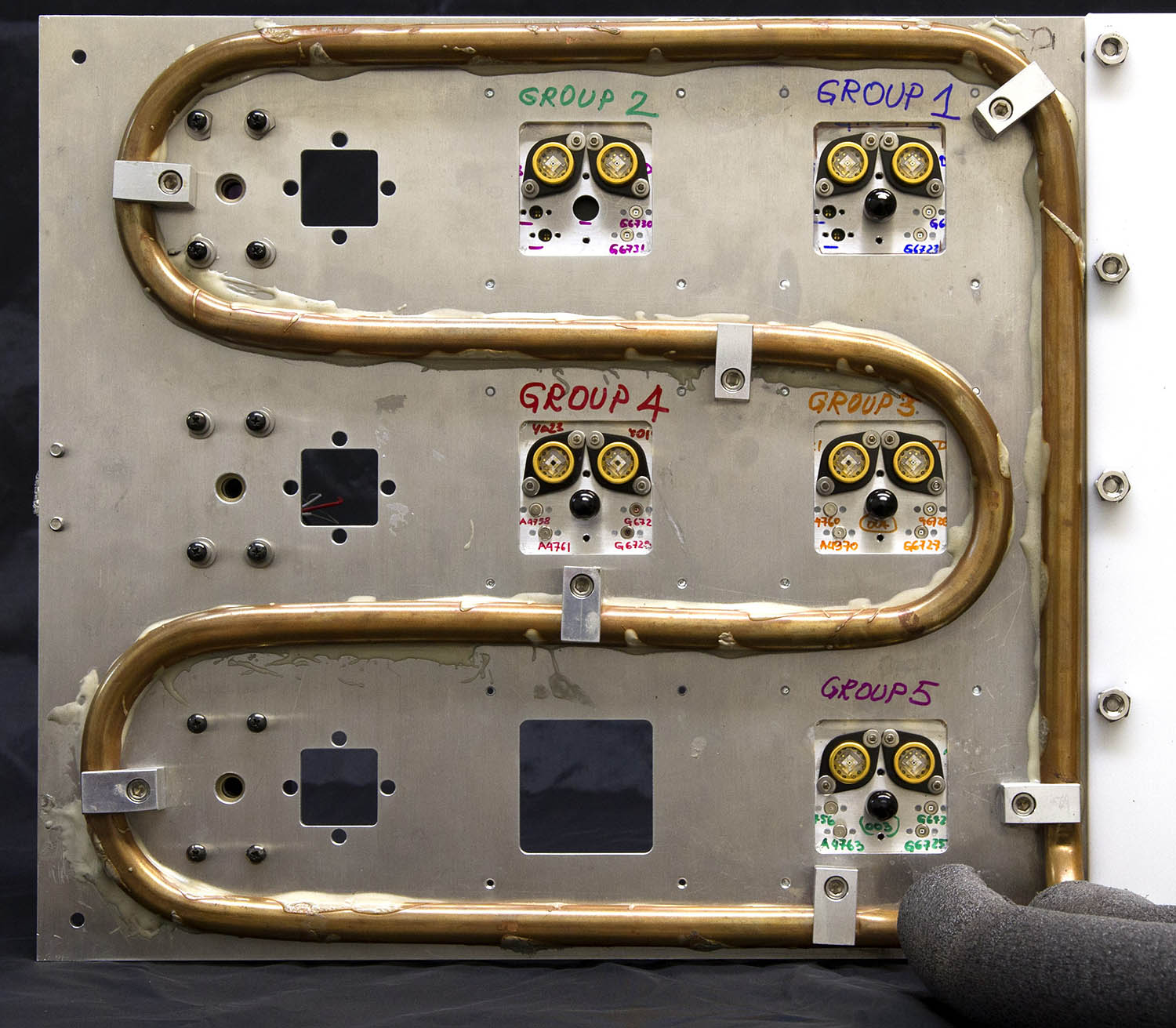}
  \caption{\label{fig:frame_plate}The main aluminum frame with all detectors groups---5 APD (right and middle column) and 3 PMT (leftmost column) groups---mounted prior to irradiation. Chilled antifreeze flowing through the copper tubing keeps the frame at $0\,\celsius$. A dry, insulating light-tight box (not shown) was placed around the frame.}
\end{figure}

For each group, each APD sample was assembled on an aluminum plate, with a PCB attached from the back (see \cref{fig:group_view}). A thermistor was attached to each plate to observe the local temperature. During irradiation, five groups of APDs and three PMTs were attached to a single aluminum frame (\cref{fig:frame_plate}) connected to an electrical ground. To suppress spontaneous thermal annealing of radiation damage during the irradiation process, the frame was cooled to ${\approx}0\,\celsius$ with chilled antifreeze pumped through a copper tube epoxied to the frame. This cooling also allowed us to conduct some testing of the APDs in situ, and observe the changing dark count rate during the irradiation process for group 5. (Without cooling, APD DCRs after irradiation could not be measured at room temperature, as our quenching circuit electronics would be saturated.)

For each of our APDs we used a passive quenching circuit with quenching resistance of $403~\kilo\ohm$, similar to that described in Ref.~\cite{cova1996} as a passive quenching circuit with current-mode output. This type of quenching circuit is appropriate for a quantum receiver satellite because of its simplicity and robustness, protecting against excessive current due to, e.g.,\ bright illumination or charged particles, or accidental high voltage spikes. Its maximum detection rate of 0.2--0.4~Mcps is lower compared to active quenching circuits, but sufficient for the detection rates expected in near-term QKD applications~\cite{bourgoin2013}. Conveniently, the long dead-time of this circuit (about $0.5$ to $1~\micro\second$) suppresses afterpulsing, even at low temperatures. Circuits for all APDs in a group were mounted on the same circuit board, outputting avalanche pulses through coaxial signal cables connected to each detector's cathode.

The breakdown voltage of each detector was found by gradually increasing the applied bias voltage until pulses due to dark counts began to appear in the trace of an oscilloscope. The oscilloscope was also used to observe the shape of the pulse at the nominal operating condition of 20~V excess bias. To determine detection performance properties, avalanche pulses were collected from each device, discriminated at 50~mV threshold and time-tagged with a resolution of 156.25~ps, while applied bias voltages and thermal parameters were simultaneously recorded at 10~Hz.

For measuring timing response properties and detection efficiency, each APD group was illuminated with a pulsing 780~nm reference laser emerging from a single-mode fiber. An optical test rig was assembled that held the optical fiber and a lens in place at ${\approx}20$~cm distance from the detector group plate. The attenuation and divergence of the laser beam was chosen such that less than one photon per pulse would be incident on each detector.

The optical test rig was placed in a cold freezer to perform low-temperature tests down to $-86\,\celsius$. The DCRs of the samples were measured either in the optical test rig with reference laser turned off, or while on the main aluminum frame within a light-tight enclosure. DCRs were averaged over several minutes (up to 15) of collected data to minimize uncertainty. Afterpulsing probability was calculated from DCR measurement data using an improved afterpulsing analysis~\cite{anisimova2017a}. For timing jitter and efficiency measurements, counts were collected for 15~minutes or until about $10^6$ detection events were registered (whichever came first).

All PMT measurements were taken while operating at $-5\,\celsius$, one of pre-set working temperatures achieved by the in-built cooler. The measurements of DCR and afterpulsing were done similarly to the APDs. For timing jitter and detection efficiency, we used a pulsed reference laser at 690~nm wavelength, with an Excelitas SLiK acting as a calibrated reference to determine the absolute efficiency.

\subsection*{Test schedule}

Prior to irradiation, we measured the breakdown voltage, DCR, efficiency, timing jitter and afterpulsing probability of all APD samples at $-20\,\celsius$. Group 4 and the control group were also characterized at lower temperatures. PMTs were tested for DCR, efficiency, timing jitter and afterpulsing probability.

At TRIUMF, each APD group (apart from the control) was in turn characterized for breakdown and DCR, then irradiated for a duration corresponding to the desired fluence for that group (actual applied fluences were within 1\% of desired, except for group 1 which received 4\% greater fluence). Immediately after irradiation the APDs were re-characterized for breakdown and DCR. These pre- and post-irradiation characterizations were performed in situ, at $0\,\celsius$, to minimize the influence of spontaneous thermal annealing. Uniquely, group 5, which received the same fluence as group 4, was held biased with its DCR recorded during the irradiation. Each PMT group was irradiated to the desired fluence, but no PMT measurements were taken in situ.

After irradiation, the APD and PMT samples were packed in a thermally isolated box filled with dry ice for transportation. This box provided temperatures no higher than $-12\,\celsius$ during the 48 hour transit. Following this, the samples were kept in a freezer at about $-20\,\celsius$ between tests. All APD samples were re-tested at $0\,\celsius$ for breakdown voltage and DCR upon arriving from the radiation facility, with no significant changes observed. PMTs were recharacterized at $-5\,\celsius$.

All APD samples were then characterized (breakdown, DCR, efficiency, jitter, and afterpulsing probability) at temperatures ranging from $-20\,\celsius$ to $-86\,\celsius$, allowing us to assess the effectiveness of cooling to mitigate damage due to irradiation. Finally, we performed thermal annealing on some groups at varying hot temperatures and durations, with further characterization at selected stages and cold temperatures.

\section*{Effects of radiation damage}
\label{sec:effects}

\begin{figure}
\includegraphics[width=0.94\columnwidth]{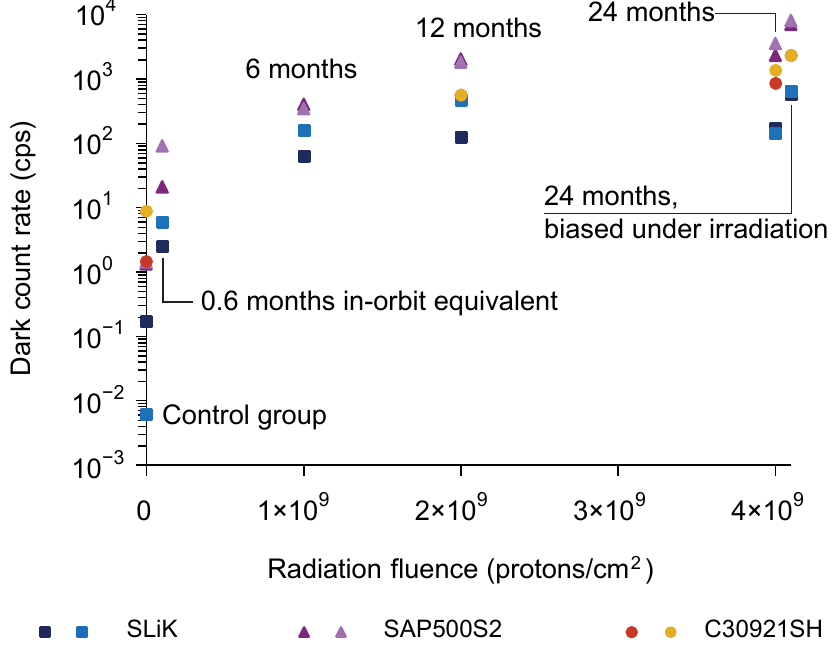}
\caption{\label{fig:darks_vs_dose}DCR of APDs after irradiation, measurement taken at $-86\,\celsius$ operation with APDs biased $20~\volt$ above their breakdown voltages. In every case, radiation damage caused a DCR increase. The APDs biased during irradiation developed a noticeably higher dark count rate.}
\end{figure}

All irradiated APDs exhibited a significant increase in their DCRs, illustrated in \cref{fig:darks_vs_dose} for $-86\,\celsius$ operating temperature, consistent with previous studies~\cite{sun2001,sun1997,tan2013}. The DCR increase in each device followed the radiation dose applied, conditional that operating temperatures were kept sufficiently low---at high temperatures, the device count rates saturated. At high doses and standard operating temperatures, the DCRs of all devices would prevent successful quantum communications---for example, Excelitas SLiK devices (overall the best performing devices) operating at $-20\,\celsius$ exhibit DCRs of the order of $10^5$~cps.

\begin{figure}
\includegraphics[width=0.94\columnwidth]{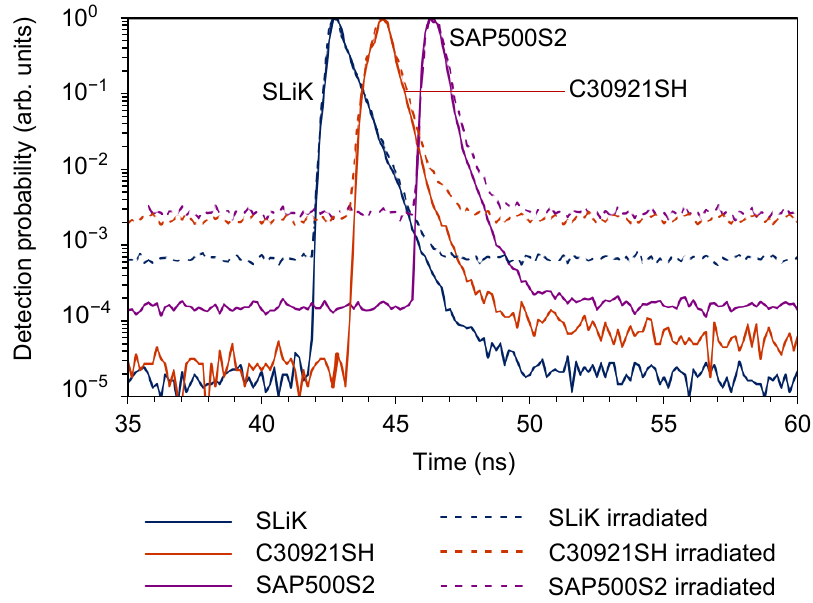}
\caption{\label{fig:jitter}Normalized timing response histogram for representative APDs from group 4 using a pulsed laser, before and after irradiation, measured at $-60\,\celsius$. The full width half maximum timing jitter before irradiation was $\approx 600\,\pico\second$ for SLiK, $\approx 550\,\pico\second$ for C30921SH, and $\approx 700\,\pico\second$ for SAP500S2. Changes in the baseline count probabilities are due to the changes in DCRs. At full width half maximum there is no noticeable change in the timing response of SLiKs and SAP500S2 before and after irradiation, and a moderate increase of $100~\pico\second$ was observed for C30921SH. Measured timing jitter includes timing jitter of the laser and time tagger.}
\end{figure}

No significant changes in breakdown voltages, pulse shapes or efficiency owing to irradiation were observed. The timing jitter of detection pulses when operating at low temperatures did not change for SLiK and SAP500S2 samples, and increased by 100~ps for C30921SH (see \cref{fig:jitter}). However, the timing jitter when operating at higher temperatures appeared to increase for all the irradiated APDs---for example, within group~4 at $-20\,\celsius$ operation, jitter increased for SLiKs by up to $80\,\pico\second$, for SAP500S2 by up to $300\,\pico\second$, and for C30921SH by up to $250\,\pico\second$. This increased timing jitter is likely due to the operation of the passive quenching mechanism at a high count rate: in this condition, avalanches often trigger before the APD voltage has fully recovered, leading to effectively lower bias voltages, which are known to have higher jitter \cite{cova1996}, for these events. Furthermore, the variation in effective bias voltages between events leads to variable current rise-times at the discriminator, and thus time-tagged events with delays dependent on the stochastic arrival of adjacent avalanches. We remark that lower jitter values than those observed in our experiment can be obtained by optimising detector electronics \cite{pugh2017,1spcmaqrhtr}.

\begin{figure}
\includegraphics[width=0.94\columnwidth]{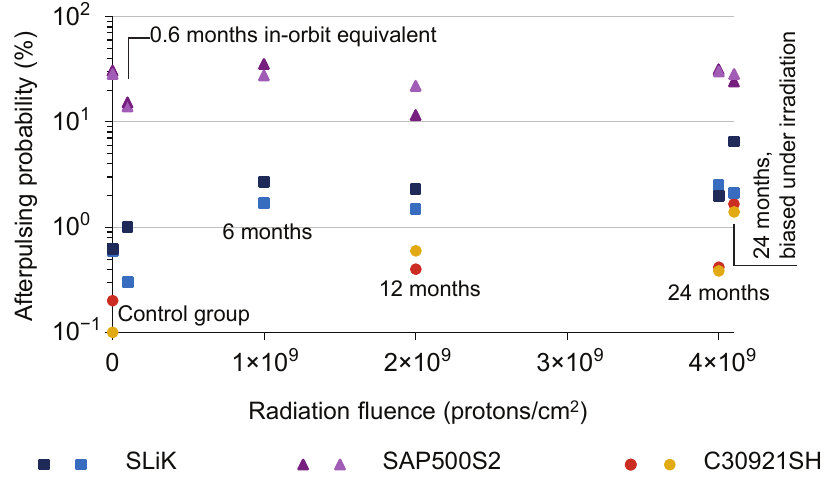}
\caption{\label{fig:afterpulses}Afterpulsing probability, measured at $-86\,\celsius$, which increased for SLiK and C30921SH devices during the first 6 to 12~month equivalent radiation dose. SAP500S2 results are high and inconsistent with respect to the applied radiation.}
\end{figure}

\begin{table*}
 \caption{\label{table:table_PMT_DCR}Four tested PMTs, their corresponding nominal fluences, equivalent to in-orbit exposures over 6, 12, and 24~months, their DCRs, afterpulsing probabilities, and jitters before and after radiation, and their detection efficiency. PMTs were not powered during the radiation. The PMT from group 9 was not irradiated, and kept as a control.\vspace{1.4mm}}
\setlength{\colwidthfluence}{2cm}
\begin{tabular}{c|c||c|c|c||c|c|c|c} \hline \hline
 \multirow{3}{*}{\textbf{Group}} & \multirow{3}{\colwidthfluence}{\textbf{Fluence @~100~MeV, protons/cm$^2$}} & \multicolumn{3}{c||}{\textbf{Before irradiation}} & \multicolumn{4}{c}{\textbf{After irradiation}} \\ \cline{3-9}
 & & \multirow{2}{*}{DRC, cps} & \multirow{2}{*}{Afterpulsing, \%} & \multirow{2}{*}{Jitter, ps} & \multirow{2}{*}{DCR, cps} & \multirow{2}{*}{Afterpulsing, \%} & \multirow{2}{*}{Jitter, ps} & \multirow{2}{*}{Efficiency, \%} \\
 & & & & & & & & \\ \hline
 6 & $10^9$ & 6.25 & 3.4 & 600 & 399 & 1.1 & 660 & 23 \\
 7 & $2\times10^9$ & 14.4 & 13.8 & 550 & 592 & 0.76 & 640 & 23 \\
 8 & $4\times10^9$ & 7 & 166 & 600 & 10 & 45 & 400 & 21 \\
 9 & 0 & 5 & 0.22 & 590 & 0.5 & 0.22 & 590 & 20 \\ \hline \hline
\end{tabular}
\end{table*}

The probability of afterpulses increased for SLiK and C30921SH samples after irradiation (\cref{fig:afterpulses}), likely due to an increased number of defects in the semiconductor crystal structure. For SAP500S2, the afterpulsing results did not show a consistent trend. Note that the afterpulsing probabilities for all SAP500S2 devices, including those in the control group, were remarkably high at lower temperatures, reaching 30\%. A longer dead-time than that provided by our circuit is clearly needed for correct operation of SAP500S2 \cite{stipcevic2013}.

APDs biased during the irradiation (group 5) developed higher DCRs than those that received the same fluence while unbiased (group 4). This result may be an important factor when planning an operational schedule for devices in an orbiting satellite---for example, it may be preferable that the detectors are off while crossing regions with higher radiation levels, such as the South Atlantic Anomaly~\cite{schaefer2016}.
\begin{figure}
\includegraphics[width=0.94\columnwidth]{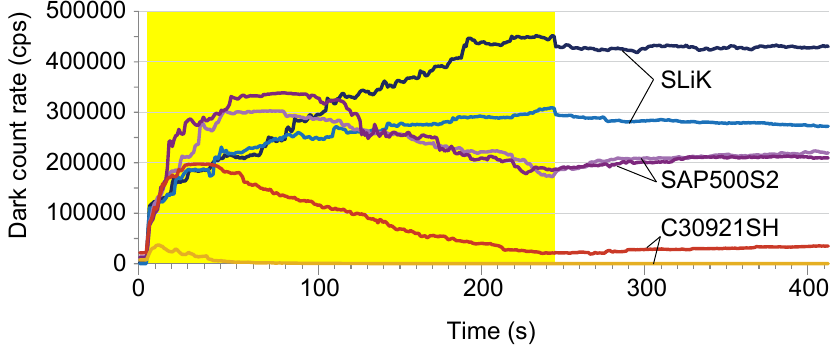}
\caption{\label{fig:darks_during_rad}DCR of APDs biased during irradiation. The highlighted portion represents the period of irradiation. While the irradiation is on, the DCR of each APD increases until saturation in the passive quenching circuit, after which saturation causes an apparent (not real) decrease in the DCRs \cite{makarov2009}. After irradiation ceased, actual DCRs slightly improved due to spontaneous annealing, leading to an apparent DCR rise in the over-saturated samples.}
\end{figure}
\cref{fig:darks_during_rad} demonstrates the dynamic change of DCRs of the APDs during irradiation. Note that all devices eventually exhibit over-saturation behaviour \cite{makarov2009} during the in-situ test.

\cref{table:table_PMT_DCR} shows the measured properties of the PMTs. In general, DCRs increased noticeably and exceeded the 200~cps desired for QKD. Anomalously, however, the PMT under the highest fluence experienced a DCR increase of merely 43\%. Given that this sample also exhibited 166\% afterpulsing probability prior to irradiation (and 45\% afterwards), it seems that the device may be defective and its properties unrepresentative. (Although, owing to a lack of time, the PMTs were not aged prior to the experiment, as is recommended by Hamamatsu. This resulted in generally elevated afterpulsing probabilities before irradiation.) DCRs as presented in \cref{table:table_PMT_DCR} were measured at 19 days after irradiation. A second DCR measurement was also performed 27 days after irradiation, where it was observed to have decreased by 10 to 25\% since the first measurement, possibly due to self-annealing, despite the PMTs being kept in a freezer at $-20\celsius$.

\section*{Mitigation of radiation damage in APD\lowercase{s}}
\label{sec:mitigation}

\subsection*{Cooling}
\begin{figure}
\includegraphics[width=0.94\columnwidth]{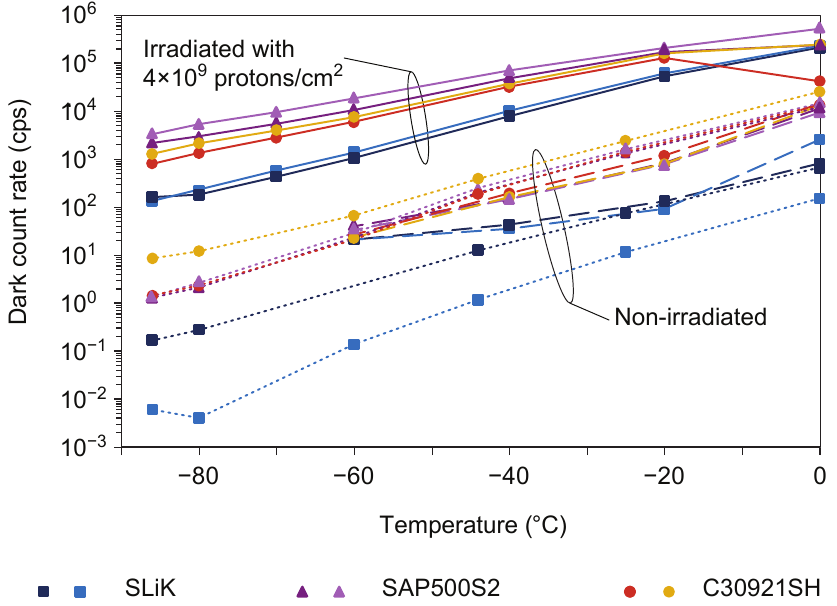}
\caption{\label{fig:darks_vs_temp}Cooling effect on DCRs of group 4 (24~month equivalent dose). Pre-irradiation data are plotted as dashed lines, post-irradiation as solid lines, and the control group as dotted lines. DCRs decrease with temperature exponentially for irradiated and non-irradiated samples.}
\end{figure}

Measurements of the detection properties of the samples reveal that radiation-induced DCRs decrease with temperature exponentially for all irradiated APDs, following the same trend as for non-radiated APDs. For SLiKs from group 4, irradiated with a 24-month-in-orbit equivalent dose, DCR dropped to 200~cps at about $-80\,\celsius$ (see \cref{fig:darks_vs_temp}).
The breakdown voltage, efficiency, and timing jitter demonstrated no significant change, though the afterpulsing probability increased significantly at lower temperatures as release time of trapped carriers extended \cite{trifonov2004}. The maximum afterpulsing probabilities in group~4 measured at $-86\,\celsius$ are 2.7\% for SLiKs, 31\% for SAP500S2, and 1.7\% for C30921S2.

Although afterpulsing is higher, we can conclude that, given sufficient cooling, SLiK SPDs can serve well for quantum protocols even after 24~months in orbit. Notably, the required temperatures are significantly above those typically reached by cryogenic coolers, and though the cooling necessary might represent a significant power demand on a small satellite system, it is nevertheless achievable. In a larger satellite or an orbital station it could be easily implemented, e.g.,\ by using solid-state thermoelectric coolers (TECs).

\subsection*{Thermal annealing}
We applied thermal annealing to all our irradiated APD samples except those in group 2 (which were set aside for laser annealing tests taking place separately~\cite{lim2017}). Samples were left at room temperature ($+20\celsius$) and in a hot-air-flow oven at $+50$, $+80$ and $+100\pm1.5\,\celsius$ for various lengths of time. After a week of annealing at room temperature there was an observed decrease of DCR, down to a factor relative to pre-annealing rates as low as 0.57 for SAP500S2 samples, and 0.71 for SLiK samples. While interesting, this rate of improvement is almost certainly too slow to be useful on a satellite platform.

\begin{figure}
\includegraphics[width=0.94\columnwidth]{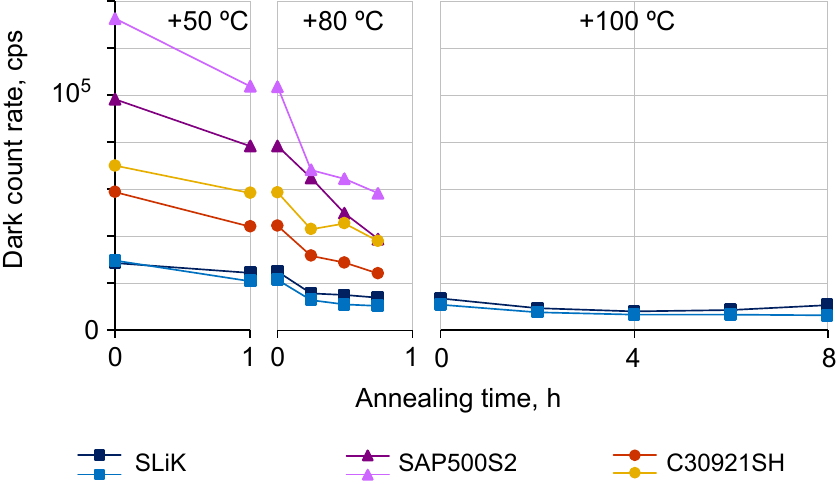}
\caption{\label{fig:annealing_gr3}DCRs measured at $-20\,\celsius$ after annealing of APDs from group 3 at $+50\,\celsius$ over 1~h, at $+80\,\celsius$ over 45~min, and for SLiK samples after further annealing at $+100\,\celsius$ over 8~h. DCRs of all APDs decrease significantly during 45~minutes of $+80\,\celsius$ annealing, and continue to decrease for a SLiK during $+100\,\celsius$ annealing, through the DCR of one of the two SLiKs increased during last 4~hours.}
\end{figure}

\begin{figure}
\includegraphics[width=0.94\columnwidth]{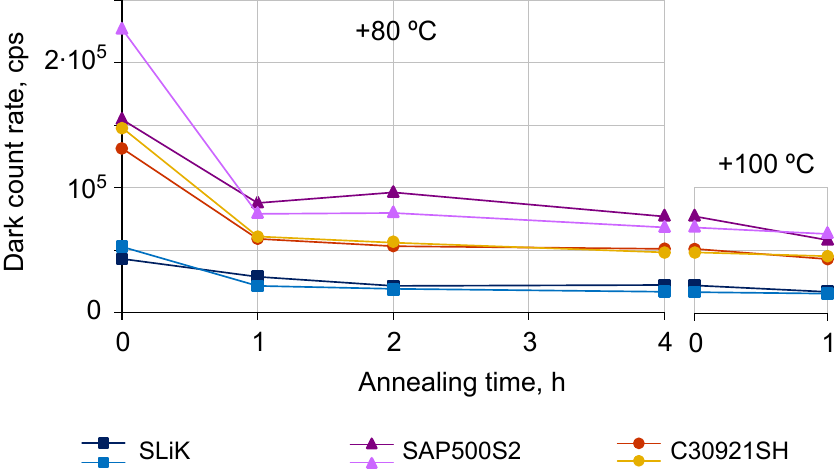}
\caption{\label{fig:annealing_gr4}DCRs measured at $-20\,\celsius$ after annealing of APDs from group 4 at $+80\,\celsius$ over 4~h, followed by annealing at $+100\,\celsius$ over 1~h. The most significant decrease of DCRs for all APDs occurred during the first hour of $+80\,\celsius$ annealing, but DCRs still continued to improve with additional annealing.}
\end{figure}

All oven-annealed APDs demonstrated more significant decreases of DCRs, with the most improvement achieving a factor 0.15 times the original pre-annealing DCR for a SLiK APD from group 3 annealed at $+50$, $+80$ and $+100\,\celsius$ (see \cref{fig:annealing_gr3})---almost a full order of magnitude DCR improvement. SAP500S2 samples saw factors as low as 0.28, and C30921SH as low as 0.3, compared to pre-annealing DCRs, both from group 4 annealed at $+80$ and $+100\,\celsius$ (see \cref{fig:annealing_gr4}).

Instead of the oven, we utilized in-built TECs for annealing of SLiKs from group 3 at $+100\,\celsius$, as this approach has the potential to simplify annealing within orbit conditions. To achieve $+100\,\celsius$ at the sensitive area while the package is at room temperature, a SLiK's TEC consumes {0.41\,\watt} of electrical power. The total annealing time with TECs was 8~h. One of the SLiKs demonstrated steady improvement of the dark count rate during that time, though the second SLiK showed some degradation after 4~h of annealing (\cref{fig:annealing_gr3}).

Breakdown voltage, detection efficiency, afterpulsing and timing response jitter of all APDs demonstrated no notable change after thermal annealing.

\section*{Conclusion}
\label{sec:conclusion}

We have conducted radiation tests of 32 APD (Excelitas and Laser Components) and 4 PMT (Hamamatsu) SPD devices, with radiation levels equivalent to lifetimes in low-Earth 600~km polar orbit of 0.6, 6, 12, and 24~months. Our performance characterization measurements showed a significant increase in DCRs for all APD devices, while there was no measurable radiation-induced degradation in the photon detection efficiency and timing jitter, and only a relatively small increase in the afterpulsing probability.

All APD samples demonstrated a significant increase in DCR due to radiation, increasing the DCR by many orders of magnitude, well above the maximum 200~cps or so required for quantum communication tasks. Subsequently, we have experimentally demonstrated that radiation damage can be successfully mitigated by sufficient cooling. For Excelitas SLiK devices, cooling to $-86\,\celsius$ was sufficient to restore the DCR to below the 200~cps level that would make quantum communications possible, even after 24-month-equivalent radiation dose.

Further DCR reduction (while preserving other performance properties) was obtained through thermal annealing. APD devices were heated at $+50$ to $+100\,\celsius$ over a few hours, in the best case resulting in a DCR only $0.15$ times that prior to annealing. It is worth noting that this approach can reduce the amount of cooling power required to reach the targeted low DCR---e.g.,\ following annealing, the SLiK APDs could achieve the target DCR of 200~cps at about $-70\,\celsius$, $16\,\celsius$ higher than prior to annealing. Thermal annealing at $+80$ to $+100\,\celsius$ seems to be the most effective, but some additional tests are required to optimize the thermal annealing for radiation damaged APDs.

Results from the PMT samples indicated small (but still significant) degradation in DCR and almost no degradation in any other measured property (efficiency, timing jitter, and afterpulsing probability) after applied radiation. This makes them a tantalizing candidate, particularly for optical inter-satellite communication applications. However, as their peak efficiency is at wavelengths where atmospheric losses are higher, they remain less interesting for ground--satellite links.

We note that, while thermal annealing is effective at reducing DCRs of APDs, the coarse method of oven-heating devices can be time and energy consuming. Alternative, more directed approaches such as the use of strong lasers may be considerably faster and require less energy (see Ref.~\citealp{lim2017}), which could be beneficial under a limited power budget of a satellite platform.

Our measurements correspond to the case where an APD is embedded on an orbiting satellite for up to two years prior to thermal annealing being applied. In a real satellite mission, thermal annealing could be applied intermittently and at regular intervals through a mission's lifetime. We speculate that doing so could repair some of the radiation-induced damage soon after it is created, thereby keeping the DCR low, delaying the necessity of deeper cooling, and extending detector lifetimes. Experimental tests of the effect of multiple irradiation and annealing cycles shall be performed. 

\begin{backmatter}

\section*{Competing interests}
  The authors declare that they have no competing interests.

\section*{Author's contributions}
T.J.\ conceived the experiment, and supervised the project with V.M. E.A.\ built the apparatus with assistance from J.-P.B. D.H.,\ L.P.P.,\ and A.S.\ performed radiation modeling. Irradiation of the samples was conducted by E.A.,\ B.L.H.,\ J.-P.B,\ L.P.P.,\ and V.M.,\ with assistance from T.J. E.A.,\ B.L.H.,\ J.-P.B.,\ and M.C.\ measured sample properties, conducted thermal annealing, and analyzed the data. E.C.,\ V.M.,\ and T.J.\ provided project management and logistical support. E.A.\ wrote the manuscript with contributions from all authors.

\section*{Acknowledgements}
This work was supported by the Canadian Space Agency, NSERC, Innovation, Science and Economic Development Canada, CFI, and Ontario MRI. We thank the staff of the Proton Irradiation Facility at TRIUMF for their professional assistance. Also we thank Ryan Cooney for helpful discussions, and Jin Gyu Lim for help with data processing.

\bibliographystyle{bmc-mathphys}


\newcommand{\BMCxmlcomment}[1]{}

\BMCxmlcomment{

<refgrp>

<bibl id="B1">
  <title><p>Overview of the {ICES}at mission</p></title>
  <aug>
    <au><snm>Schutz</snm><fnm>B. E.</fnm></au>
    <au><snm>Zwally</snm><fnm>H. J.</fnm></au>
    <au><snm>Shuman</snm><fnm>C. A.</fnm></au>
    <au><snm>Hancock</snm><fnm>D.</fnm></au>
    <au><snm>DiMarzio</snm><fnm>J. P.</fnm></au>
  </aug>
  <source>Geophys. Res. Lett.</source>
  <pubdate>2005</pubdate>
  <volume>32</volume>
  <note>{page L21S01}</note>
</bibl>

<bibl id="B2">
  <title><p>Space-qualified silicon avalanche-photodiode single-photon-counting
  modules</p></title>
  <aug>
    <au><snm>Sun</snm><fnm>X</fnm></au>
    <au><snm>Krainak</snm><fnm>MA</fnm></au>
    <au><snm>Abshire</snm><fnm>JB</fnm></au>
    <au><snm>Spinhirne</snm><fnm>JD</fnm></au>
    <au><snm>Trottier</snm><fnm>C</fnm></au>
    <au><snm>Davies</snm><fnm>M</fnm></au>
    <au><snm>Dautet</snm><fnm>H</fnm></au>
    <au><snm>Allan</snm><fnm>GR</fnm></au>
    <au><snm>Lukemire</snm><fnm>AT</fnm></au>
    <au><snm>Vandiver</snm><fnm>JC</fnm></au>
  </aug>
  <source>J. Mod. Optic</source>
  <pubdate>2004</pubdate>
  <volume>51</volume>
  <fpage>1333</fpage>
</bibl>

<bibl id="B3">
  <title><p>Scintillation Detectors for Charged Particles and
  Photons</p></title>
  <aug>
    <au><snm>Lecoq</snm><fnm>P.</fnm></au>
  </aug>
  <source>Detectors for Particles and Radiation. Part 1: Principles and
  Methods</source>
  <publisher>Berlin, Heidelberg: Springer Berlin Heidelberg</publisher>
  <editor>Fabjan, C. W. and Schopper, H.</editor>
  <pubdate>2011</pubdate>
  <fpage>45</fpage>
  <lpage>-71</lpage>
</bibl>

<bibl id="B4">
  <title><p>Photon counting module for laser time transfer via {E}arth orbiting
  satellite</p></title>
  <aug>
    <au><snm>Prochazka</snm><fnm>I</fnm></au>
    <au><snm>Yang</snm><fnm>F</fnm></au>
  </aug>
  <source>J. Mod. Opt.</source>
  <pubdate>2009</pubdate>
  <volume>56</volume>
  <issue>2-3</issue>
  <fpage>253</fpage>
  <lpage>260</lpage>
</bibl>

<bibl id="B5">
  <title><p>Ground to satellite secure key exchange using quantum
  cryptography</p></title>
  <aug>
    <au><snm>Rarity</snm><fnm>J. G.</fnm></au>
    <au><snm>Owens</snm><fnm>P. C. M.</fnm></au>
    <au><snm>Tapster</snm><fnm>P. R.</fnm></au>
  </aug>
  <source>New J. Phys.</source>
  <pubdate>2002</pubdate>
  <volume>4</volume>
</bibl>

<bibl id="B6">
  <title><p>Space-quest, experiments with quantum entanglement in
  space</p></title>
  <aug>
    <au><snm>Ursin</snm><fnm>R.</fnm></au>
    <au><snm>Jennewein</snm><fnm>T.</fnm></au>
    <au><snm>Kofler</snm><fnm>J.</fnm></au>
    <au><snm>Perdigues</snm><fnm>J. M.</fnm></au>
    <au><snm>Cacciapuoti</snm><fnm>L.</fnm></au>
    <au><snm>Matos</snm><fnm>C. J.</fnm></au>
    <au><snm>Aspelmeyer</snm><fnm>M.</fnm></au>
    <au><snm>Valencia</snm><fnm>A.</fnm></au>
    <au><snm>Scheidl</snm><fnm>T.</fnm></au>
    <au><snm>Acin</snm><fnm>A.</fnm></au>
    <au><snm>Barbieri</snm><fnm>C.</fnm></au>
    <au><snm>Bianco</snm><fnm>G.</fnm></au>
    <au><snm>Brukner</snm><fnm>C.</fnm></au>
    <au><snm>Capmany</snm><fnm>J.</fnm></au>
    <au><snm>Cova</snm><fnm>S.</fnm></au>
    <au><snm>Giggenbach</snm><fnm>D.</fnm></au>
    <au><snm>Leeb</snm><fnm>W.</fnm></au>
    <au><snm>Hadfield</snm><fnm>R. H.</fnm></au>
    <au><snm>Laflamme</snm><fnm>R.</fnm></au>
    <au><snm>L{\"u}tkenhaus</snm><fnm>N.</fnm></au>
    <au><snm>Milburn</snm><fnm>G.</fnm></au>
    <au><snm>Peev</snm><fnm>M.</fnm></au>
    <au><snm>Ralph</snm><fnm>T.</fnm></au>
    <au><snm>Rarity</snm><fnm>J.</fnm></au>
    <au><snm>Renner</snm><fnm>R.</fnm></au>
    <au><snm>Samain</snm><fnm>E.</fnm></au>
    <au><snm>Solomos</snm><fnm>N.</fnm></au>
    <au><snm>Tittel</snm><fnm>W.</fnm></au>
    <au><snm>Torres</snm><fnm>J. P.</fnm></au>
    <au><snm>Toyoshima</snm><fnm>M.</fnm></au>
    <au><snm>Ortigosa Blanch</snm><fnm>A.</fnm></au>
    <au><snm>Pruneri</snm><fnm>V.</fnm></au>
    <au><snm>Villoresi</snm><fnm>P.</fnm></au>
    <au><snm>Walmsley</snm><fnm>I.</fnm></au>
    <au><snm>Weihs</snm><fnm>G.</fnm></au>
    <au><snm>Weinfurter</snm><fnm>H.</fnm></au>
    <au><snm>Zukowski</snm><fnm>M.</fnm></au>
    <au><snm>Zeilinger</snm><fnm>A.</fnm></au>
  </aug>
  <source>Europhys. News</source>
  <pubdate>2009</pubdate>
  <volume>40</volume>
  <issue>3</issue>
  <fpage>26</fpage>
  <lpage>29</lpage>
</bibl>

<bibl id="B7">
  <title><p>Feasibility of satellite quantum key distribution</p></title>
  <aug>
    <au><snm>Bonato</snm><fnm>C</fnm></au>
    <au><snm>Tomaello</snm><fnm>A</fnm></au>
    <au><snm>Deppo</snm><fnm>VD</fnm></au>
    <au><snm>Naletto</snm><fnm>G</fnm></au>
    <au><snm>Villoresi</snm><fnm>P</fnm></au>
  </aug>
  <source>New J. Phys.</source>
  <pubdate>2009</pubdate>
  <volume>11</volume>
  <issue>4</issue>
  <fpage>045017</fpage>
</bibl>

<bibl id="B8">
  <title><p>A comprehensive design and performance analysis of low {E}arth
  orbit satellite quantum communication</p></title>
  <aug>
    <au><snm>Bourgoin</snm><fnm>J. P.</fnm></au>
    <au><snm>Meyer Scott</snm><fnm>E.</fnm></au>
    <au><snm>Higgins</snm><fnm>B. L.</fnm></au>
    <au><snm>Helou</snm><fnm>B.</fnm></au>
    <au><snm>Erven</snm><fnm>C.</fnm></au>
    <au><snm>H{\" u}bel</snm><fnm>H.</fnm></au>
    <au><snm>Kumar</snm><fnm>B.</fnm></au>
    <au><snm>Hudson</snm><fnm>D.</fnm></au>
    <au><snm>D'Souza</snm><fnm>I.</fnm></au>
    <au><snm>Girard</snm><fnm>R.</fnm></au>
    <au><snm>Laflamme</snm><fnm>R.</fnm></au>
    <au><snm>Jennewein</snm><fnm>T.</fnm></au>
  </aug>
  <source>New J. Phys.</source>
  <pubdate>2013</pubdate>
  <volume>15</volume>
  <fpage>023006</fpage>
</bibl>

<bibl id="B9">
  <title><p>Chinese satellite is one giant step for the quantum
  internet</p></title>
  <aug>
    <au><snm>Gibney</snm><fnm>E.</fnm></au>
  </aug>
  <source>Nature</source>
  <pubdate>2016</pubdate>
  <volume>535</volume>
  <fpage>478</fpage>
  <lpage>479</lpage>
</bibl>

<bibl id="B10">
  <title><p>Space {QUEST} mission proposal: experimentally testing decoherence
  due to gravity</p></title>
  <aug>
    <au><snm>Joshi</snm><fnm>S. K.</fnm></au>
    <au><snm>Pienaar</snm><fnm>J.</fnm></au>
    <au><snm>Ralph</snm><fnm>T. C.</fnm></au>
    <au><snm>Cacciapuoti</snm><fnm>L.</fnm></au>
    <au><snm>McCutcheon</snm><fnm>W.</fnm></au>
    <au><snm>Rarity</snm><fnm>J.</fnm></au>
    <au><snm>Giggenbach</snm><fnm>D.</fnm></au>
    <au><snm>Makarov</snm><fnm>V.</fnm></au>
    <au><snm>Fuentes</snm><fnm>I.</fnm></au>
    <au><snm>Scheidl</snm><fnm>T.</fnm></au>
    <au><snm>Beckert</snm><fnm>E.</fnm></au>
    <au><snm>Bourennane</snm><fnm>M.</fnm></au>
    <au><snm>Bruschi</snm><fnm>D. E.</fnm></au>
    <au><snm>Cabello</snm><fnm>A.</fnm></au>
    <au><snm>Capmany</snm><fnm>J.</fnm></au>
    <au><snm>Carrasco</snm><fnm>J. A.</fnm></au>
    <au><snm>Carrasco Casado</snm><fnm>A.</fnm></au>
    <au><snm>Diamanti</snm><fnm>E.</fnm></au>
    <au><snm>Du{\v s}ek</snm><fnm>M.</fnm></au>
    <au><snm>Elser</snm><fnm>D.</fnm></au>
    <au><snm>Gulinatti</snm><fnm>A.</fnm></au>
    <au><snm>Hadfield</snm><fnm>R. H.</fnm></au>
    <au><snm>Jennewein</snm><fnm>T.</fnm></au>
    <au><snm>Kaltenbaek</snm><fnm>R.</fnm></au>
    <au><snm>Krainak</snm><fnm>M. A.</fnm></au>
    <au><snm>Lo</snm><fnm>H. K.</fnm></au>
    <au><snm>Marquardt</snm><fnm>C</fnm></au>
    <au><snm>Mataloni</snm><fnm>P.</fnm></au>
    <au><snm>Milburn</snm><fnm>G.</fnm></au>
    <au><snm>Peev</snm><fnm>M.</fnm></au>
    <au><snm>Poppe</snm><fnm>A.</fnm></au>
    <au><snm>Pruneri</snm><fnm>V.</fnm></au>
    <au><snm>Renner</snm><fnm>R.</fnm></au>
    <au><snm>Salomon</snm><fnm>C.</fnm></au>
    <au><snm>Skaar</snm><fnm>J.</fnm></au>
    <au><snm>Solomos</snm><fnm>N.</fnm></au>
    <au><snm>Stip{\v c}evi{\' c}</snm><fnm>M.</fnm></au>
    <au><snm>Torres</snm><fnm>J. P.</fnm></au>
    <au><snm>Toyoshima</snm><fnm>M.</fnm></au>
    <au><snm>Villoresi</snm><fnm>P.</fnm></au>
    <au><snm>Walmsley</snm><fnm>I.</fnm></au>
    <au><snm>Weihs</snm><fnm>G.</fnm></au>
    <au><snm>Weinfurter</snm><fnm>H.</fnm></au>
    <au><snm>Zeilinger</snm><fnm>A.</fnm></au>
    <au><snm>{\. Z}ukowski</snm><fnm>M.</fnm></au>
    <au><snm>Ursin</snm><fnm>R.</fnm></au>
  </aug>
</bibl>

<bibl id="B11">
  <title><p>Experimental quantum key distribution with simulated
  ground-to-satellite photon losses and processing limitations</p></title>
  <aug>
    <au><snm>Bourgoin</snm><fnm>JP</fnm></au>
    <au><snm>Gigov</snm><fnm>N</fnm></au>
    <au><snm>Higgins</snm><fnm>BL</fnm></au>
    <au><snm>Yan</snm><fnm>Z</fnm></au>
    <au><snm>Meyer Scott</snm><fnm>E</fnm></au>
    <au><snm>Khandani</snm><fnm>AK</fnm></au>
    <au><snm>L\"utkenhaus</snm><fnm>N</fnm></au>
    <au><snm>Jennewein</snm><fnm>T</fnm></au>
  </aug>
  <source>Phys. Rev. A</source>
  <publisher>American Physical Society</publisher>
  <pubdate>2015</pubdate>
  <volume>92</volume>
  <fpage>052339</fpage>
</bibl>

<bibl id="B12">
  <title><p>Single-photon detectors for optical quantum information
  applications</p></title>
  <aug>
    <au><snm>Hadfield</snm><fnm>R. H.</fnm></au>
  </aug>
  <source>Nat. Photonics</source>
  <pubdate>2009</pubdate>
  <volume>3</volume>
  <fpage>696</fpage>
  <lpage>-705</lpage>
</bibl>

<bibl id="B13">
  <title><p>Measurement of proton radiation damage to {Si} avalanche
  photodiodes</p></title>
  <aug>
    <au><snm>Sun</snm><fnm>X</fnm></au>
    <au><snm>Reusser</snm><fnm>D</fnm></au>
    <au><snm>Dautet</snm><fnm>H</fnm></au>
    <au><snm>Abshire</snm><fnm>JB</fnm></au>
  </aug>
  <source>IEEE Trans. Electron. Devices</source>
  <pubdate>1997</pubdate>
  <volume>44</volume>
  <fpage>2160</fpage>
</bibl>

<bibl id="B14">
  <title><p>Proton radiation damage of {Si} {APD} single photon
  counters</p></title>
  <aug>
    <au><snm>Sun</snm><fnm>X</fnm></au>
    <au><snm>Dautet</snm><fnm>H</fnm></au>
  </aug>
  <source>Proc.\ of IEEE Radiation Effects Data Workshop 2001</source>
  <pubdate>2001</pubdate>
  <fpage>146</fpage>
</bibl>

<bibl id="B15">
  <title><p>Silicon avalanche photodiode operation and lifetime analysis for
  small satellites</p></title>
  <aug>
    <au><snm>Tan</snm><fnm>YC</fnm></au>
    <au><snm>Chandrasekara</snm><fnm>R</fnm></au>
    <au><snm>Cheng</snm><fnm>C</fnm></au>
    <au><snm>Ling</snm><fnm>A</fnm></au>
  </aug>
  <source>Opt. Express</source>
  <pubdate>2013</pubdate>
  <volume>21</volume>
  <fpage>16946</fpage>
</bibl>

<bibl id="B16">
  <title><p>Single photon avalanche diodes for space applications</p></title>
  <aug>
    <au><snm>Marisaldi</snm><fnm>M.</fnm></au>
    <au><snm>Maccagnani</snm><fnm>P.</fnm></au>
    <au><snm>Moscatelli</snm><fnm>F.</fnm></au>
    <au><snm>Labanti</snm><fnm>C.</fnm></au>
    <au><snm>Fuschino</snm><fnm>F.</fnm></au>
    <au><snm>Prest</snm><fnm>M.</fnm></au>
    <au><snm>Berra</snm><fnm>A.</fnm></au>
    <au><snm>Bolognini</snm><fnm>D.</fnm></au>
    <au><snm>Ghioni</snm><fnm>M.</fnm></au>
    <au><snm>Rech</snm><fnm>I.</fnm></au>
    <au><snm>Gulinatti</snm><fnm>A.</fnm></au>
    <au><snm>Giudice</snm><fnm>A.</fnm></au>
    <au><snm>Simmerle</snm><fnm>G.</fnm></au>
    <au><snm>Rubini</snm><fnm>D.</fnm></au>
    <au><snm>Candelori</snm><fnm>A.</fnm></au>
    <au><snm>Mattiazzo</snm><fnm>S.</fnm></au>
  </aug>
  <source>Nuclear Science Symposium and Medical Imaging Conference (NSS/MIC),
  2011 IEEE</source>
  <pubdate>2011</pubdate>
  <fpage>129</fpage>
  <lpage>134</lpage>
</bibl>

<bibl id="B17">
  <title><p>Radiation tests of single photon avalanche diode for space
  applications</p></title>
  <aug>
    <au><snm>Moscatelli</snm><fnm>F</fnm></au>
    <au><snm>Marisaldi</snm><fnm>M</fnm></au>
    <au><snm>Maccagnani</snm><fnm>P</fnm></au>
    <au><snm>Labanti</snm><fnm>C</fnm></au>
    <au><snm>Fuschino</snm><fnm>F</fnm></au>
    <au><snm>Prest</snm><fnm>M</fnm></au>
    <au><snm>Berra</snm><fnm>A</fnm></au>
    <au><snm>Bolognini</snm><fnm>D</fnm></au>
    <au><snm>Ghioni</snm><fnm>M</fnm></au>
    <au><snm>Rech</snm><fnm>I</fnm></au>
    <au><snm>Gulinatti</snm><fnm>A</fnm></au>
    <au><snm>Giudice</snm><fnm>A</fnm></au>
    <au><snm>Simmerle</snm><fnm>G</fnm></au>
    <au><snm>Candelori</snm><fnm>A</fnm></au>
    <au><snm>Mattiazzo</snm><fnm>S</fnm></au>
    <au><snm>Sun</snm><fnm>X</fnm></au>
    <au><snm>Cavanaugh</snm><fnm>JF</fnm></au>
    <au><snm>Rubini</snm><fnm>D</fnm></au>
  </aug>
  <source>Nucl. Instr. Meth. Phys. Res. A</source>
  <pubdate>2013</pubdate>
  <volume>711</volume>
  <fpage>65</fpage>
  <lpage>72</lpage>
</bibl>

<bibl id="B18">
  <title><p>Ultra-low noise single-photon detector based on Si avalanche
  photodiode</p></title>
  <aug>
    <au><snm>Kim</snm><fnm>Y. S.</fnm></au>
    <au><snm>Jeong</snm><fnm>Y. C.</fnm></au>
    <au><snm>Sauge</snm><fnm>S.</fnm></au>
    <au><snm>Makarov</snm><fnm>V.</fnm></au>
    <au><snm>Kim</snm><fnm>Y. H.</fnm></au>
  </aug>
  <source>Rev. Sci. Instrum.</source>
  <pubdate>2011</pubdate>
  <volume>82</volume>
  <fpage>093110</fpage>
</bibl>

<bibl id="B19">
  <title><p>Low-noise single-photon detectors for long-distance free-space
  quantum communication</p></title>
  <aug>
    <au><snm>Anisimova</snm><fnm>E.</fnm></au>
    <au><snm>Nikulov</snm><fnm>D.</fnm></au>
    <au><snm>Hu</snm><fnm>S. S.</fnm></au>
    <au><snm>Bourgon</snm><fnm>M.</fnm></au>
    <au><snm>Ursin</snm><fnm>R.</fnm></au>
    <au><snm>Jennewein</snm><fnm>T.</fnm></au>
    <au><snm>Makarov</snm><fnm>V.</fnm></au>
  </aug>
  <note>{i}n preparation; preliminary results presented at QCrypt 2015 in
  Tokyo, Japan and Single Photon Workshop 2015 in Geneva, Switzerland</note>
</bibl>

<bibl id="B20">
  <title><p>The effects of radiation on electronic systems</p></title>
  <aug>
    <au><snm>Messenger</snm><fnm>GC</fnm></au>
    <au><snm>Ash</snm><fnm>MS</fnm></au>
  </aug>
  <publisher>Van Nostrand Reinhold Co. Inc., New York, NY</publisher>
  <pubdate>1986</pubdate>
</bibl>

<bibl id="B21">
  <title><p>Radiation effects on microelectronics in space</p></title>
  <aug>
    <au><snm>Srour</snm><fnm>J.R.</fnm></au>
    <au><snm>McGarrity</snm><fnm>J.M.</fnm></au>
  </aug>
  <source>Proc. IEEE</source>
  <pubdate>1988</pubdate>
  <volume>76</volume>
  <issue>11</issue>
  <fpage>1443</fpage>
  <lpage>1469</lpage>
</bibl>

<bibl id="B22">
  <title><p>Radiation damage of electronic and optoelectronic devices in
  space</p></title>
  <aug>
    <au><snm>Johnston</snm><fnm>A. H.</fnm></au>
  </aug>
  <source>The 4th Int. Workshop on Radiation Effects on Semiconductor Devices
  for Space Application</source>
  <pubdate>2000</pubdate>
</bibl>

<bibl id="B23">
  <title><p>Universal Damage Factor for Radiation-Induced Dark Current in
  Silicon Devices</p></title>
  <aug>
    <au><snm>Srour</snm><fnm>J. R.</fnm></au>
    <au><snm>Lo</snm><fnm>D. H.</fnm></au>
  </aug>
  <source>IEEE Trans. Nucl. Sci.</source>
  <pubdate>2000</pubdate>
  <volume>47</volume>
  <issue>6</issue>
  <fpage>2451</fpage>
  <lpage>-2459</lpage>
</bibl>

<bibl id="B24">
  <title><p>A comparison of {M}onte {C}arlo and analytic treatments of
  displacement damage in {Si} microvolumes</p></title>
  <aug>
    <au><snm>Dale</snm><fnm>C. J.</fnm></au>
    <au><snm>Chen</snm><fnm>L.</fnm></au>
    <au><snm>McNulty</snm><fnm>P. J.</fnm></au>
    <au><snm>Marshall</snm><fnm>P. W.</fnm></au>
    <au><snm>Burke</snm><fnm>E. A.</fnm></au>
  </aug>
  <source>IEEE Trans. Nucl. Sci.</source>
  <pubdate>1994</pubdate>
  <volume>41</volume>
  <issue>6</issue>
  <fpage>1974</fpage>
  <lpage>1983</lpage>
</bibl>

<bibl id="B25">
  <title><p>The Influence of Structural Characteristics on the Response of
  Silicon Avalanche Photodiodes to Proton Irradiation</p></title>
  <aug>
    <au><snm>Becker</snm><fnm>HN</fnm></au>
    <au><snm>Miyahira</snm><fnm>TF</fnm></au>
    <au><snm>Johnston</snm><fnm>AH</fnm></au>
  </aug>
  <source>IEEE Trans. Nucl. Sci.</source>
  <pubdate>2003</pubdate>
  <volume>50</volume>
  <issue>6</issue>
  <fpage>1974</fpage>
  <lpage>-1981</lpage>
</bibl>

<bibl id="B26">
  <title><p>Avalanche photodiodes and quenching circuits for single-photon
  detection</p></title>
  <aug>
    <au><snm>Cova</snm><fnm>S.</fnm></au>
    <au><snm>Ghioni</snm><fnm>M.</fnm></au>
    <au><snm>Lacaita</snm><fnm>A.</fnm></au>
    <au><snm>Samori</snm><fnm>C.</fnm></au>
    <au><snm>Zappa</snm><fnm>F.</fnm></au>
  </aug>
  <source>Appl. Opt.</source>
  <publisher>OSA</publisher>
  <pubdate>1996</pubdate>
  <volume>35</volume>
  <issue>12</issue>
  <fpage>1956</fpage>
  <lpage>-1976</lpage>
</bibl>

<bibl id="B27">
  <title><p>Airborne demonstration of a quantum key distribution receiver
  payload</p></title>
  <aug>
    <au><snm>Pugh</snm><fnm>C. J.</fnm></au>
    <au><snm>Kaiser</snm><fnm>S.</fnm></au>
    <au><snm>Bourgoin</snm><fnm>J. P.</fnm></au>
    <au><snm>Jin</snm><fnm>J.</fnm></au>
    <au><snm>Sultana</snm><fnm>N.</fnm></au>
    <au><snm>Agne</snm><fnm>S.</fnm></au>
    <au><snm>Anisimova</snm><fnm>E.</fnm></au>
    <au><snm>Makarov</snm><fnm>V.</fnm></au>
    <au><snm>Choi</snm><fnm>E.</fnm></au>
    <au><snm>Higgins</snm><fnm>B. L.</fnm></au>
    <au><snm>Jennewein</snm><fnm>T.</fnm></au>
  </aug>
  <note>{Q}uantum {S}ci.\ {T}echnol.\ (in press)</note>
</bibl>

<bibl id="B28">
  <note>{E}xcelitas SPCM-AQRH-TR timing resolution optimised single photon
  counting module,
  \url{http://www.excelitas.com/Downloads/DTS_SPCM-AQRH-TR.pdf}, visited 15 May
  2017</note>
</bibl>

<bibl id="B29">
  <title><p>Characterization of a commercially available large area, high
  detection efficiency single-photon avalanche diode</p></title>
  <aug>
    <au><snm>Stip{\v c}evi{\' c}</snm><fnm>M.</fnm></au>
    <au><snm>Wang</snm><fnm>D.</fnm></au>
    <au><snm>Ursin</snm><fnm>R.</fnm></au>
  </aug>
  <source>J. Lightwave Technol.</source>
  <pubdate>2013</pubdate>
  <volume>31</volume>
  <issue>23</issue>
  <fpage>3591</fpage>
  <lpage>3596</lpage>
</bibl>

<bibl id="B30">
  <title><p>Observation and modeling of the {S}outh {A}tlantic {A}nomaly in low
  {E}arth orbit using photometric instrument data</p></title>
  <aug>
    <au><snm>Schaefer</snm><fnm>R. K.</fnm></au>
    <au><snm>Paxton</snm><fnm>L. J.</fnm></au>
    <au><snm>Selby</snm><fnm>C.</fnm></au>
    <au><snm>Ogorzalek</snm><fnm>B.</fnm></au>
    <au><snm>Romeo</snm><fnm>G.</fnm></au>
    <au><snm>Wolven</snm><fnm>B.</fnm></au>
    <au><snm>Hsieh</snm><fnm>S. Y.</fnm></au>
  </aug>
  <source>Space Weather</source>
  <pubdate>2016</pubdate>
  <volume>14</volume>
  <issue>5</issue>
  <fpage>330</fpage>
  <lpage>342</lpage>
</bibl>

<bibl id="B31">
  <title><p>Controlling passively quenched single photon detectors by bright
  light</p></title>
  <aug>
    <au><snm>Makarov</snm><fnm>V.</fnm></au>
  </aug>
  <source>New J. Phys.</source>
  <pubdate>2009</pubdate>
  <volume>11</volume>
  <issue>6</issue>
  <fpage>065003</fpage>
</bibl>

<bibl id="B32">
  <title><p>Single photon counting at telecom wavelength and quantum key
  distribution</p></title>
  <aug>
    <au><snm>Trifonov</snm><fnm>A</fnm></au>
    <au><snm>Subacius</snm><fnm>D</fnm></au>
    <au><snm>Berzanskis</snm><fnm>A</fnm></au>
    <au><snm>Zavriyev</snm><fnm>A</fnm></au>
  </aug>
  <source>J. Mod. Opt.</source>
  <pubdate>2004</pubdate>
  <volume>51</volume>
  <issue>9-10</issue>
  <fpage>1399</fpage>
</bibl>

<bibl id="B33">
  <title><p>Laser annealing heals radiation damage in single-photon avalanche
  photodiodes</p></title>
  <aug>
    <au><snm>Lim</snm><fnm>JG</fnm></au>
    <au><snm>Anisimova</snm><fnm>E</fnm></au>
    <au><snm>Higgins</snm><fnm>BL</fnm></au>
    <au><snm>Bourgoin</snm><fnm>JP</fnm></au>
    <au><snm>Jennewein</snm><fnm>T</fnm></au>
    <au><snm>Makarov</snm><fnm>V</fnm></au>
  </aug>
  <source>EPJ Quantum Technol.</source>
  <volume>4</volume>
  <fpage>11</fpage>
</bibl>

</refgrp>
} 

\end{backmatter}
\end{document}